\documentclass[twocolumn,3p,11pt,number]{elsarticle}
\usepackage{graphicx}
\usepackage{lineno,hyperref}
\usepackage{amsmath}
\modulolinenumbers[5]
\usepackage{lineno}

\journal{Nuclear Instruments and Methods in Physics Research A}

\begin{document}

\begin{frontmatter}

\title{Study of timing characteristics of a 3 m long plastic scintillator counter using waveform digitizers}

\author[1]{A.~Blondel}
\author[2]{D.~Breton}
\author[1]{A.~Dubreuil}
\author[3]{A.~Khotyantsev}
\author[1]{A.~Korzenev\corref{cor1}}
\author[2]{J.~Maalmi}
\author[3]{A.~Mefodev}
\author[1]{P.~Mermod\corref{cor2}}
\author[1]{E.~Noah}

\cortext[cor1]{E-mail: alexander.korzenev@cern.ch}
\cortext[cor2]{E-mail: philippe.mermod@cern.ch}

\address[1]{DPNC, Section de Physique, Universit\'{e} de Gen\`{e}ve, Geneva, Switzerland}
\address[2]{Laboratoire de L'acc\'{e}l\'{e}rateur Lin\'{e}aire from CNRS/IN2P3, Centre scientifique d'Orsay, France}
\address[3]{Institute for Nuclear Research of the Russian Academy of Sciences, Moscow, Russia}

\begin{abstract}
A plastic scintillator bar with dimensions \mbox{300~cm $\times$ 2.5~cm $\times$ 11~cm} was exposed to a focused muon beam to study its light yield and timing characteristics as a function of position and angle of incidence. The scintillating light was read out at both ends by photomultiplier tubes whose pulse shapes were recorded by waveform digitizers. Results obtained with the WAVECATCHER and SAMPIC digitizers are analyzed and compared. A 
 discussion of the various factors affecting the timing resolution is presented. Prospects for applications of plastic scintillator technology in large-scale particle physics detectors with timing resolution around 100~ps are provided in light of the results.
\end{abstract}


\end{frontmatter}



\section{Introduction}
\label{sec:Introduction}

Plastic scintillator detectors have been extensively used in particle physics experiments for decades. In large-scale experiments, they 
are typically arranged as an array
covering a large surface which can provide a fast trigger signal or particle identification using the time-of-flight (ToF) technique. 
Depending on the bar dimensions, scintillator type and light readout sensor, the time resolution\footnote{In this paper, the time resolution is expressed as the rms value if not otherwise specified.} for such detectors typically ranges from 50~ps (0.5~m bars of the ToF system of MICE~\cite{Bertoni:2010by}) to 350~ps (6.8~m bars of the ToF system of OPAL~\cite{Ahmet:1990eg}).


In practice, bars which are made of a bulk scintillator do not exceed 3~m in length. This restriction comes naturally from light attenuation within the plastic and an uncertainty related to the dispersion of photon path lengths which becomes dominant for long bars. 
Moreover, this uncertainty grows exponentially with
decreasing bar thickness \cite{Perrino:1996pr}.
It makes a bar cross section close to a square shape advantageous in detectors~\cite{Tsujita:1996tk,Denisov:2004ag,Kichimi:2000,Wu:2005xk}. However, when a detector covers a large surface, for reasons of economy, the bar thickness along the beam is often chosen to be significantly smaller 
than its width. In this case the thickness becomes a limiting factor for the precision of the time measurement. 
Recent examples of detectors using this type of bars
are the trigger hodoscopes system in COMPASS~\cite{Bernet:2005yy} and in the NA61/SHINE ToF detector~\cite{Abgrall:2014xwa}. 

Another example of a detector which combines the requirements of a large covered surface and an excellent time resolution is the timing detector of the proposed SHiP experiment at the CERN SPS~\cite{Bonivento:2013jag}. To efficiently distinguish between vertices from random muon crossings and genuine particle decays, the SHiP timing detector needs to cover a 6~m $\times$ 12~m area with a time resolution of 100~ps or better~\cite{Anelli:2015pba} at an affordable price, which is a challenge. One option considered in the SHiP technical proposal is an array of 3~m long plastic scintillator bars 
with the light collected
by photomultiplier tubes (PMTs)~\cite{Anelli:2015pba}. Another feature of SHiP is a software trigger running on an online computer farm, thus favoring the use of a DAQ electronics which has the particularity to tolerate relatively high event rates and at the same time allow for each channel to operate in a self-triggering mode. 


Novel types of acquisition electronics which perform  waveform sampling using a switched capacitor array (SCA) 
have only recently been employed
in particle physics experiments
\cite{Delagnes:WAVECATCHER,Delagnes:SAMPIC}.
The use of an analogue memory which is added in parallel with a delay line allows
for analog signal sampling
at a very high rate.
In addition, having the waveform recorded, one can extract various kinds of information 
such as
baseline, amplitude, charge and time. The measurements presented in this article with a 3~m bar were made with the two acquisition modules WAVECATCHER \cite{Delagnes:WAVECATCHER} and SAMPIC \cite{Delagnes:SAMPIC}. 
The latter is proposed for the data acquisition system for the SHiP timing detector.
The test-bench used here can thus be considered as a prototype for the 
design of the timing detector of the SHiP experiment described in the technical proposal~\cite{Anelli:2015pba}.

The article is organized as follows.
The experimental setup is described in Section~\ref{sec:Setup}.
Section~\ref{DAQ} provides a detailed description of the DAQ electronics.
The analysis procedure is presented in Section~\ref{Analysis}.
The results of the measurements are discussed in Section~\ref{sec:Results}.
Finally, a summary is given in Section~\ref{sec:Conclusions}.

\section{Experimental setup}
\label{sec:Setup}

\begin{figure}
\includegraphics[width=\linewidth]{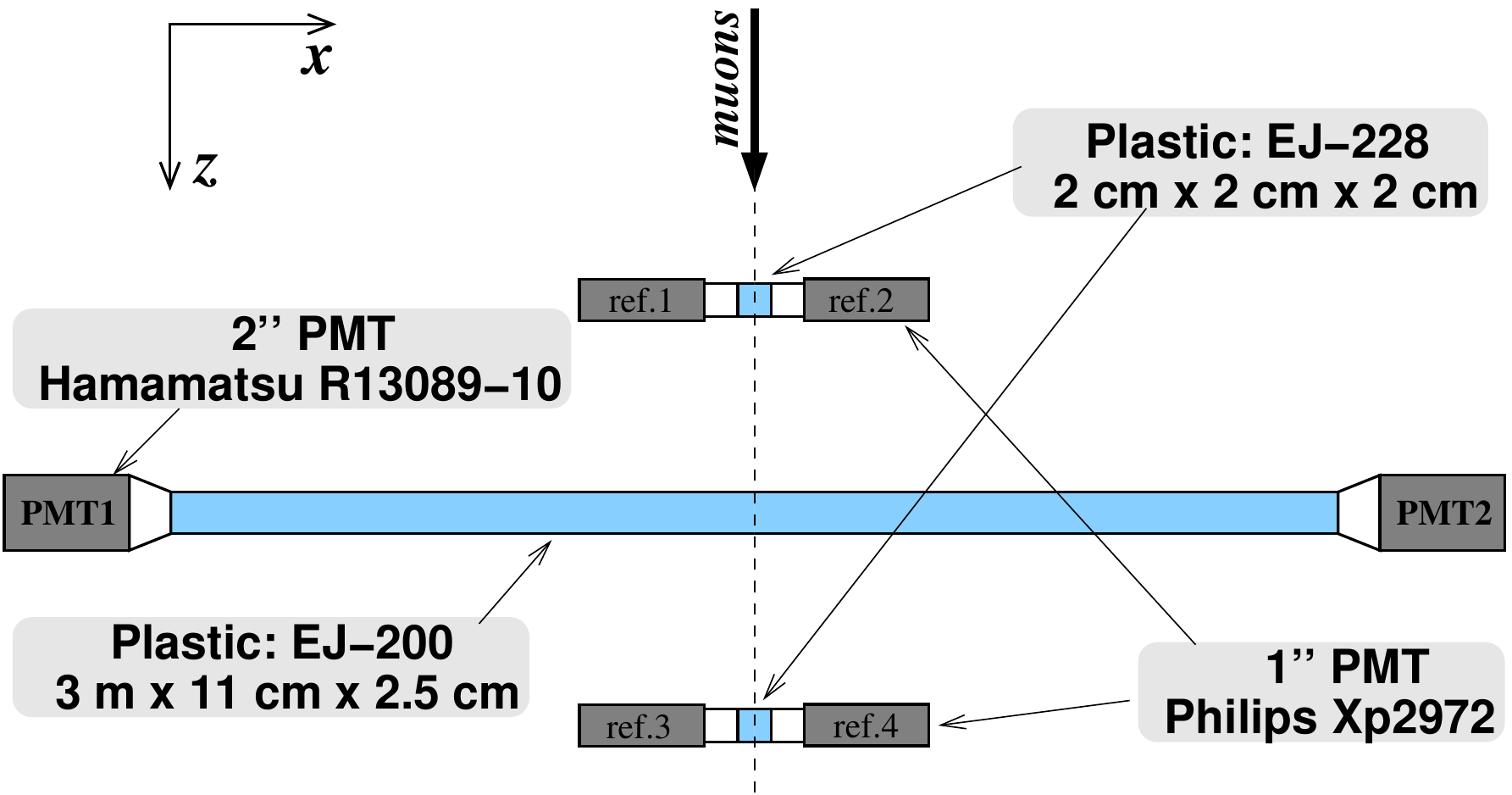}
\caption{Schematic top view of the experimental setup. The outputs of all 6 PMTs are connected to a single acquisition module.}
\label{fig:setup}
\end{figure}

We present results of test-beam measurements which took place at the CERN PS in June 2016. The layout of the setup is shown in Fig.~\ref{fig:setup}. The coordinate system is chosen such that the $z$ axis is directed along the beam, the $x$ axis is along the bar and the $y$ axis is directed vertically in such a way that the coordinate system is right-handed. The origin of the system is at the left side of the bar, in the center of the $yz$ cross section.

\subsection{Plastic bar and PMTs}
\label{sec:Setup_plastic}

\begin{figure}
\centering
\includegraphics[width=0.95\linewidth]{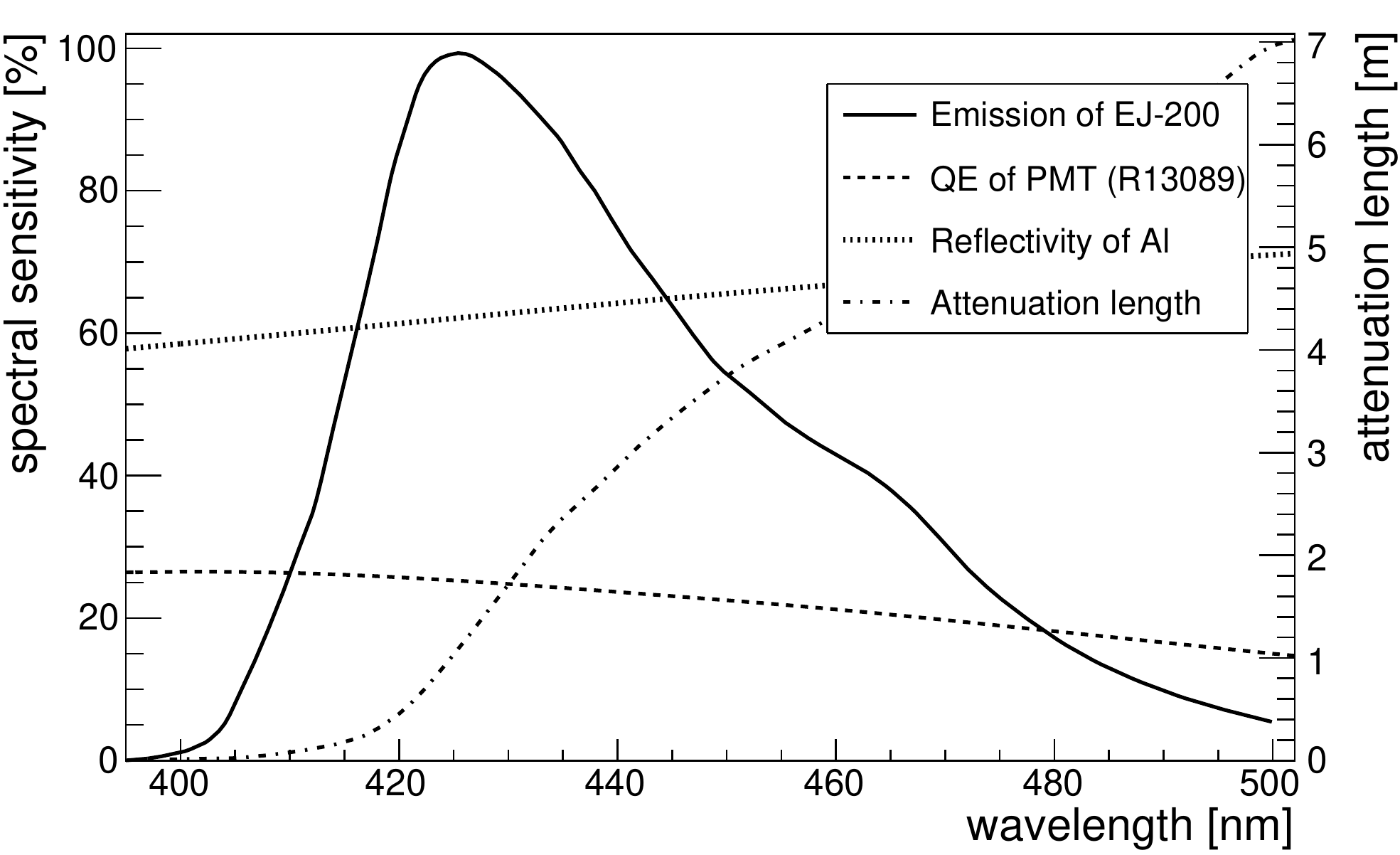}
\caption{
  Emission spectrum of EJ-200~\cite{SCIONIX} (arbitrary units) overlaid with the quantum efficiency of the PMT~\cite{Hamamatsu}, the reflection efficiency of an aluminum foil~\cite{Wu:2005xk} and the attenuation length\protect\footnotemark~\cite{Brooks_CLASS} (right axis).
}
\label{fig:sensitivity}
\end{figure}

The scintillator bar was purchased from the SCIONIX Radiation Detector
\& Crystals company~\cite{SCIONIX}. The bar length is 300~cm and its transverse cross section is 2.5~cm $\times$ 11~cm. The two larger surfaces of the bar (300~cm $\times$ 11~cm) were in contact with a casting form and had no other preparation. The four other surfaces were diamond milled. The choice of plastic was primarily driven by the length of the bar: EJ-200 provides an optimal combination of a suitable optical attenuation length, fast timing and high light output. The properties of EJ-200 quoted by the producer are: a rise time of 0.9~ns; a decay time of 2.1~ns; a bulk attenuation length of 4~m; and a refraction index of 1.58. The peak in the emission spectrum resides in the violet region of the visible spectrum. As shown in Fig.~\ref{fig:sensitivity}, this spectrum is compatible with the sensitivity region of the PMT and the reflection efficiency of an aluminum foil which was used to wrap the bar.

\footnotetext{Measurements presented in Ref.\,\cite{Brooks_CLASS} are done for BC-412 which is based on polyvinyltoluene as EJ-200.}

The bar is attached via tapered light guides to 2'' phototubes on both ends. The fast Hamamatsu R13089-10 PMT~\cite{Hamamatsu} is chosen because of its good time resolution and moderate cost.  It has a linear-focused dynode structure with 8 stages and a typical anode gain value of $3.2 \times 10^5$. 
The voltage divider optimized for timing applications was provided by the company.
The PMT output signal was coupled directly to the acquisition module.
This results in a signal amplitude 
in the range $30-150$~mV
(given for the most probable value) 
which fits perfectly the dynamic range of the acquisition modules.
The quantum efficiency of the photocathode as given by the manufacturer is 25\% at the 
emission peak for
the scintillator (see Fig.~\ref{fig:sensitivity}). Parameters relevant for the precision time measurements are a rise time of 2~ns and a transit time spread\footnote{A spread of fluctuations of the transit time for a single photoelectron~\cite{Hamamatsu}.} of 230~ps.

The phototubes were pressed towards the light guides. 
The probable presence of air gaps between the photocathode and plastic
may however reduce the amount of photons at large angles due to total internal reflection. Also, the cross-sectional area of the bar is larger than the area of the  photocathode by about 34\%. Due to phase-space conservation of the photon flux the light output should be reduced by about the same amount in the case where an interaction took place in the proximity of the PMT. 

The bar and PMTs were fixed to an aluminum frame which could be moved vertically and horizontally with respect to the beam.

\subsection{Beam and trigger system}

Measurements were carried out using a 10~GeV/$c$ muon beam produced by interactions of 24~GeV/$c$ protons from the CERN PS accelerator with closed shutters at the T9 beam line of the East Hall.

The trigger was formed by the coincidence of signals from two beam counters installed 50~cm up- and downstream with respect to the bar under test as shown in Fig.~\ref{fig:setup}. The counters are shaped as cubes with 2~cm sides made of a fast EJ-228 scintillator with rise and decay constants 0.5~ns and 1.4~ns, respectively. They were coupled to 1'' PMTs (Philips Xp2972) from two sides via 5~cm long light guides.

The trigger time is calculated as an average of the measurements of all four trigger PMTs. This time is used as a reference for the measurement of the counter under test. The resolution of the trigger system is derived from the width of a distribution of the time difference between the time measurements by up- and downstream counters. It is found to be 40~ps. 
Another contribution to the trigger time resolution is associated with a finite size of the beam counters. It was estimated to be 36~ps assuming a uniform distribution of the beam within the counter area. Both contributions are further subtracted in quadrature from the uncertainty obtained with the counter under test.

\section{DAQ electronics}
\label{DAQ}

The major design criterion for the DAQ system is an internal time resolution which has to be much better than the expected resolution of the scintillator counter. The chosen electronics modules 
WAVECATCHER~\cite{Delagnes:WAVECATCHER} and SAMPIC~\cite{Delagnes:SAMPIC} are based on waveform digitizer ASICs which have been developed by LAL and IRFU teams since 1992. The technology employs a circular buffer, based on arrays of switched capacitors (SCA) which record an analogue signal at very high rate. The time information is extracted with an interpolation of samples in the leading edge of the signal which permits reaching a timing accuracy of about 5~ps\footnote{The value indicates the measured resolution of the ASICs. It was obtained by splitting a generated pulse between two channels and delaying one of them~\cite{Delagnes:WAVECATCHER,Delagnes:SAMPIC}. The resolution is defined as the measured jitter on the time difference between the two pulses.}. Fast detectors for ToF measurements thus represent a natural target for these digitizers.

In both modules, common clock and trigger signals can be provided externally. Besides that, each channel also integrates a discriminator that can trigger individually or participate in a more complex trigger. This capability makes these digitizer modules very suitable for a test bench experiment
since no extra electronics are required for the trigger. The digitizers combine in a single module the functions of TDC and ADC, and allow to perform a digital constant fraction discrimination (dCFD)
which, in the case of the WAVECATCHER, is a part of the firmware.

The desktop versions of the acquisition modules were used. They house the USB2 interface which permits a connection to PC with 480~Mbit/s. The acquisition software runs under Windows saving data directly on disk.

The analogue signal sampling in WAVECATCHER is based on the SAMLONG SCA 
\cite{Delagnes:WAVECATCHER}. The chip was designed with the AMS CMOS 0.35-$\mu$m technology
and houses two fully differential channels. The bandwidth of 500~MHz is well suited for detection of very short pulses. An analogue signal is sampled at a rate which can be set between 0.4 and 3.2~GS/s.
Voltages stored in capacitors are further digitized with commercial ADCs at a much lower rate (10 to 20~MHz). This results in an overall readout dead time of about 120~$\mu$s for the full sampling depth of the circular buffer which comprises 1024 cells. However one can define for the readout an interval of interest which can be a subset of the whole buffer. The board is DC-coupled with a dynamic range of 2.5~V and adjustable offsets are coded over 16 bits. In the acquisition of data presented here the desktop module with 8 channels running at 3.2~GS/s (buffer depth 320 ns) has been used.

The SAMPIC acquisition module is based on the cognominal ASIC designed to be the first TDC directly working on analogue signals \cite{Delagnes:SAMPIC}. SAMPIC was developed in the frame of the R\&D project aiming at a Waveform and Time to Digital Converter (WTDC) multi-channel chip. It was initially intended to address needs for high precision timing detectors (5~ps) required by ATLAS AFP and SuperB FTOF. 

SAMPIC is a 16-channel ASIC designed with the AMS CMOS 0.18-$\mu$m technology. Each channel associates a DLL-based TDC providing a raw time with a 64-cell ultra-fast analogue memory sampling up to \mbox{10.2~GS/s} thus assuring high-resolution timing information. Analogue data are digitized by an on-chip ADC (8 to 11 bits). The relatively small sampling depth allows for a low dead time, around 1.5~$\mu$s when using the ADC in the 11-bit mode and of the order of 0.2~$\mu$s in the 8-bit mode.  The chip is designed to offer a signal bandwidth of \mbox{1.5~GHz} and a usable dynamic range of 1~V.

In the work presented here SAMPIC was exploited at the 3.2~GS/s sampling rate. The typical width of a signal in the scintillator counter is 25~ns. Thus, SAMPIC is generally not able to record the full signal waveform, but the sampling depth of 20~ns is enough to determine the baseline and cover the front edge of the signal. The sampling rate can in principle be lowered to 1.6~GS/s which would increase the time window by a factor of two.

Each SAMPIC channel integrates a discriminator which, contrary to WAVECATCHER, can trigger itself independently of other channels. This mode was used in the data taking. Coincidence of reference counters has been checked offline and selects about 1\% of recorded events. The rest of the data represent interactions from muon pile-up and cosmics.

\begin{figure}[t]
\centering
\includegraphics[width=0.99\linewidth]{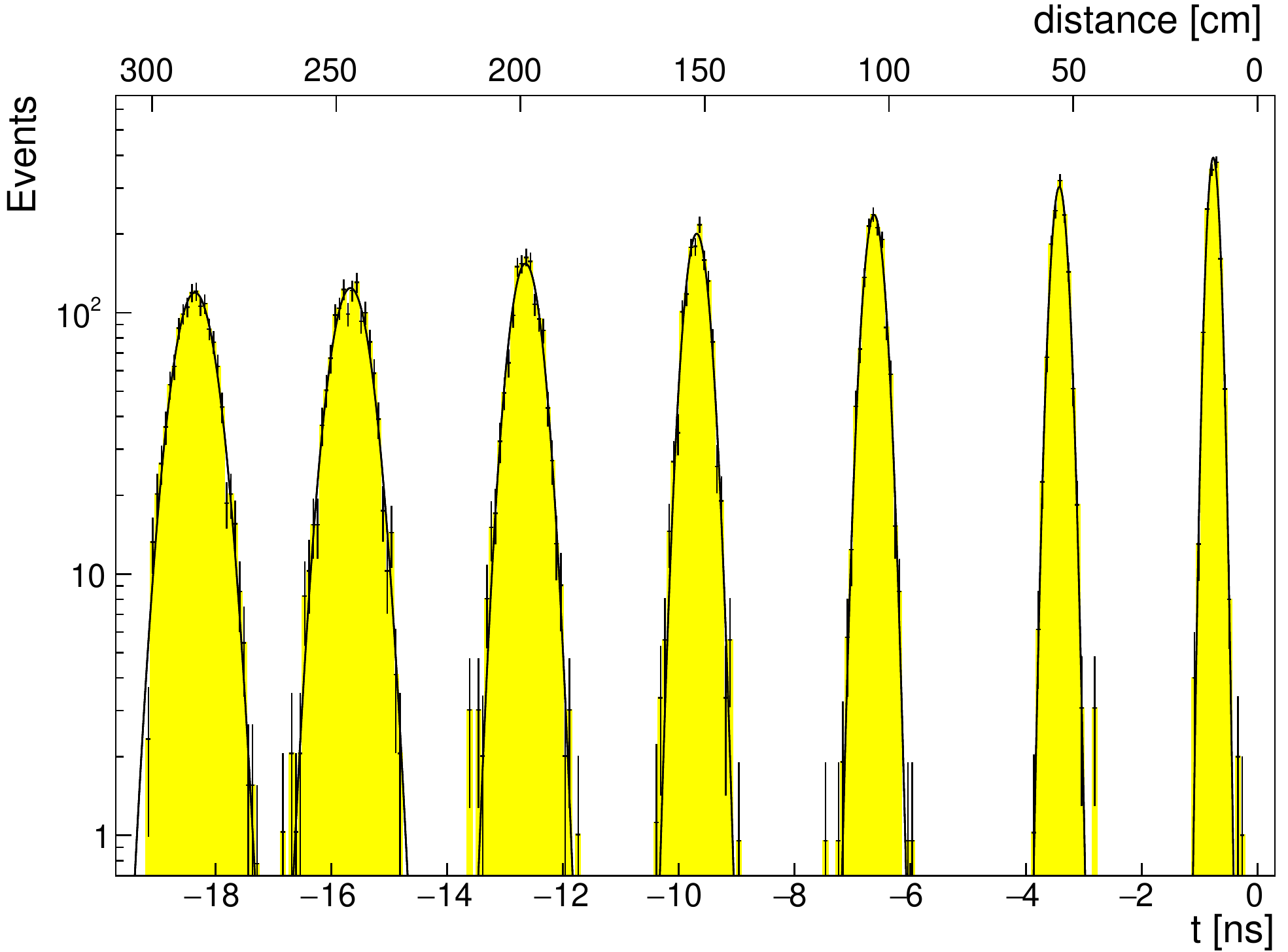}
\caption{
  Superposition of distributions of time registered by PMT1
  for different positions of an interaction point along the $x$ axis.
}
\label{fig:DT}
\end{figure}

\section{Analysis}
\label{Analysis}

The time resolution of the counter is studied using either WAVECATCHER or SAMPIC as a function of the position of a charged-particle interaction 
either along $x$ or $y$ axis
and angle of incidence of the particle trajectory.
The counter was exposed to the muon beam to collect about two thousand triggers for every point, after which the bar was manually either shifted or rotated to the next position. Measurements were done for 15 points along the $x$ axis of the bar\footnote{Two extra points at $x=2$~cm and $x=298$~cm at $y=0$~cm were taken at the end of the run with WAVECATCHER.}. 
For each of these, two measurements were made: one at the center $y=0$~cm and another at the edge $y=4$~cm. Rotations were performed only for a beam position at the center of the bar, with 8 different angles between 40$^{\circ}$ and 90$^{\circ}$ in the horizontal $xz$ plane and three different angles between 60$^{\circ}$ and 90$^{\circ}$ in the vertical $yz$ plane.  

\begin{figure}
\centering
\includegraphics[width=0.99\linewidth]{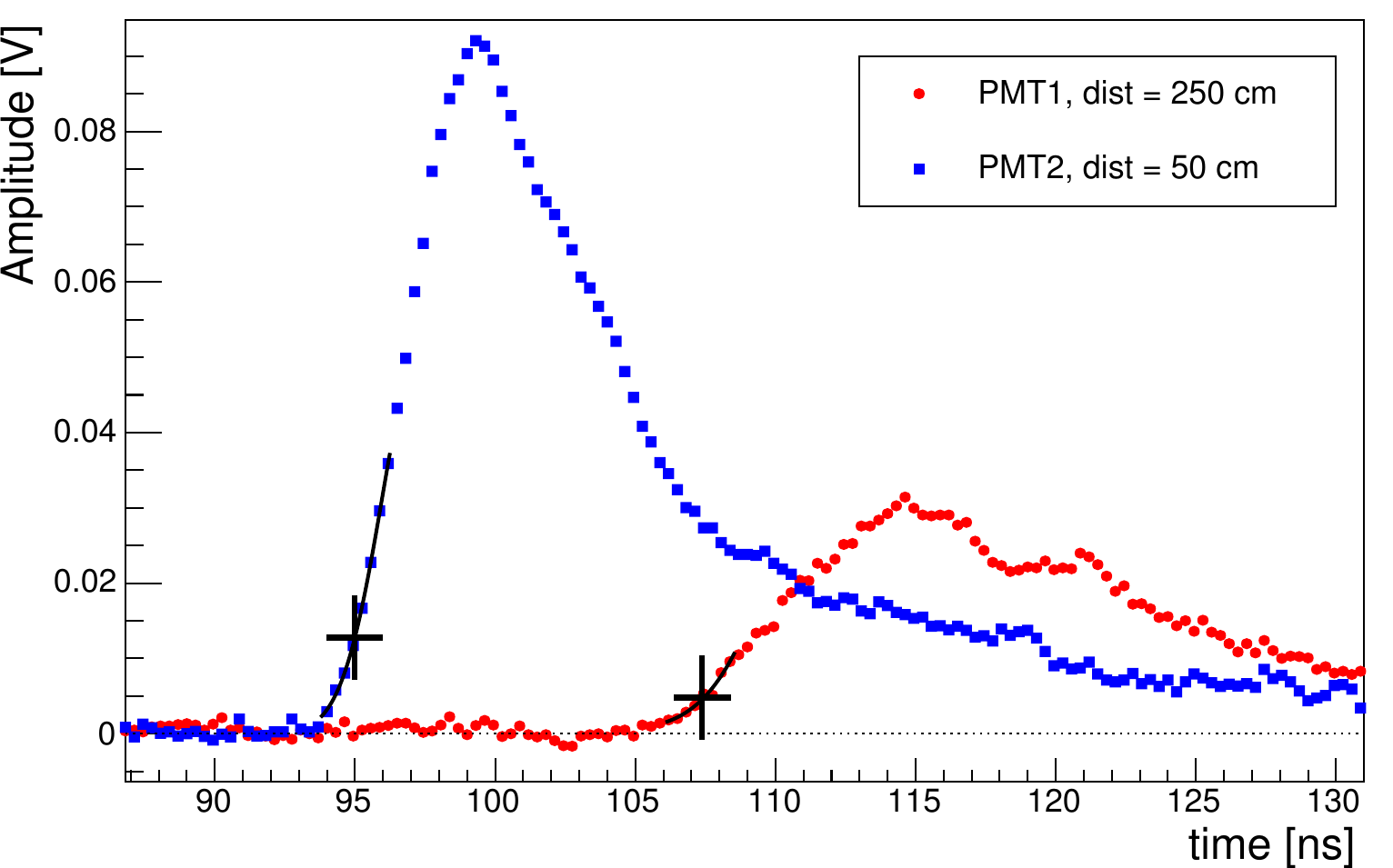}
\caption{Typical waveforms of a signal detected by PMT1 and PMT2
  and recorded by WAVECATCHER at $x=250$~cm. Crosses show signal times at a fraction 0.14 of the pulse amplitudes.}
\label{fig:wf}
\end{figure}

In the analysis, the time $t$ corresponding to either PMT1 or PMT2 is the measured time subtracted  from the reference time (defined as the mean value of the times registered by the two trigger counters). Examples of time spectra as measured for different positions along $x$ are shown in Fig.~\ref{fig:DT}. The time spectra can be reasonably approximated by Gaussian functions. 
For each position, the variance and the mean of the function were used to obtain the time resolution and the peak position of the distribution.

There are several techniques for the extraction of time from the sampled waveforms~\cite{Nelson}. However it was shown that results obtained with a digital constant fraction discrimination technique typically show a better time resolution~\cite{Wang:2015rva,Delagnes:2016hdo}. In the dCFD approach the signal time is determined by the crossing point of the interpolated digitized signal at the threshold which, in turn, is a constant fraction of the pulse amplitude.

The variance of time measurements can be expressed as~\cite{Delagnes:2016hdo}
\begin{equation}
\sigma_t^2 = \Big( \frac{\sigma_u}{ du/dt} \Big)^2 + Const ,
\label{eq:dT}
\end{equation}
where $\sigma_u$ is a variance of the measured voltage and $du/dt$ is a slope of the rising edge of a signal. $Const$ is an uncertainty due to the limited precision of measurements (for instance the TDC jitter) which is independent of a signal.

Waveforms of typical signals detected by PMT1 and PMT2 at $x=250$~cm and recorded by the WAVECATCHER are shown in Fig.~\ref{fig:wf}. Two methods are applied to determine the crossing point of dCFD. In the first method a linear interpolation between two neighboring samples is utilized. In the second method a Gaussian function is used to approximate a part of the waveform which includes eight consecutive samples. 
The second method can give an advantage in the case of a smaller number of detected photons
resulting from 
long signal propagation along the bar.
Indeed, if $\sigma_u$ is dominated by statistical fluctuations the fit can improve the precision. 
On the other hand, if the signal propagation distance is short and, in turn, the statistical contribution to $\sigma_u$ becomes less significant, the correlation effect between samples~\cite{Delagnes:2016hdo} negates the improvement due to the fit, leading to no significant difference between the two methods.

\begin{figure}
\centering
\includegraphics[width=0.99\linewidth]{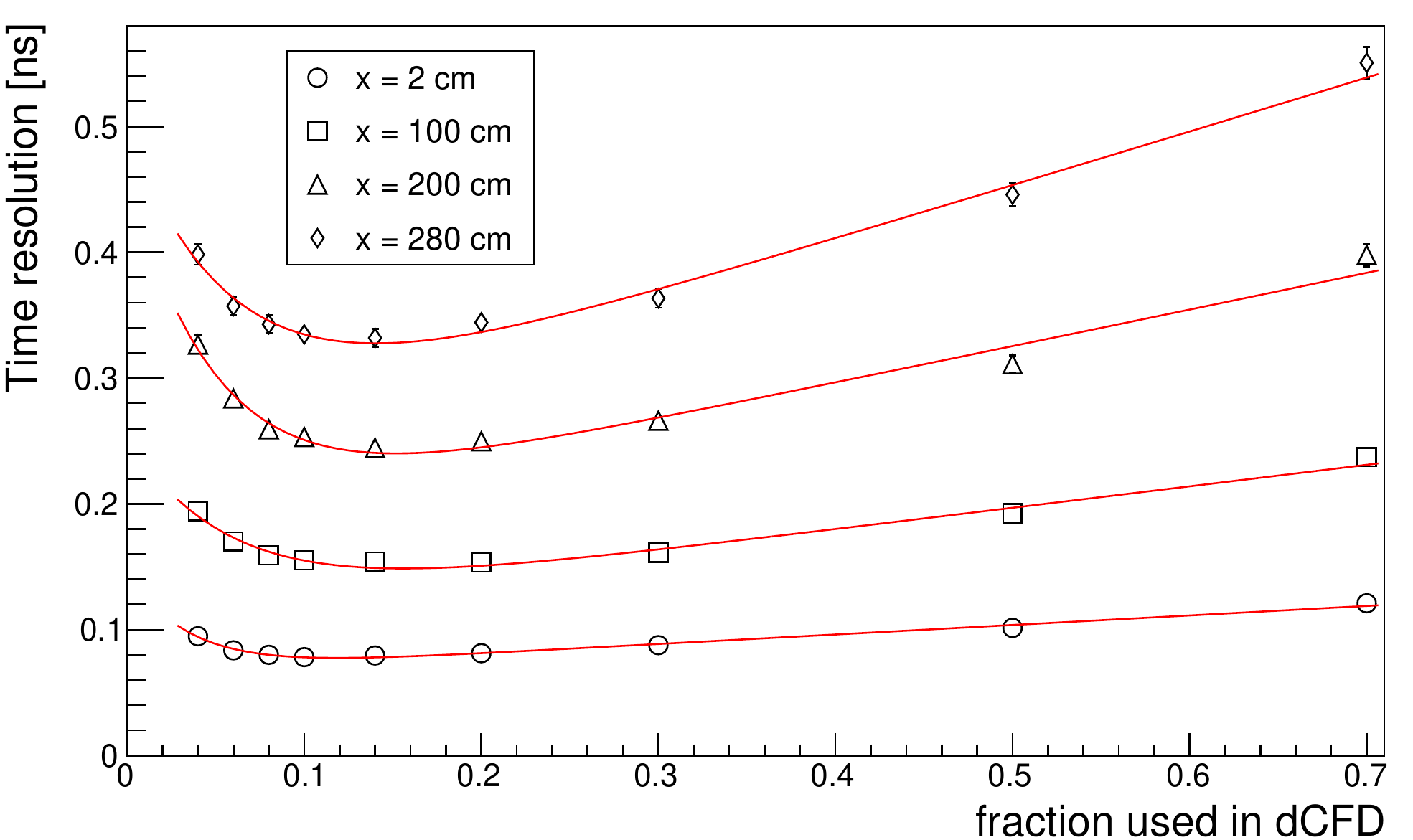}
\caption{Time resolution of the counter as viewed by PMT1 as a function of the fraction 
   used in dCFD. The different sets of points correspond to different distances
   between the interaction position and the PMT.}
\label{fig:CFD}
\end{figure}

The value of the dCFD fraction should be chosen as a compromise between growing fluctuations $\sigma_u$ of the rising amplitude and an improving steepness of the slope $du/dx$ (see Eq.~\ref{eq:dT}).
A scan is performed in the interval 0.04 $-$ 0.7 to determine the optimal fraction.
The time resolution of the counter as viewed by PMT1 as a function of the dCFD fraction is shown on Fig.~\ref{fig:CFD}. The optimal value depends weakly on the distance between the interaction point and the PMT. For the analysis presented in this work 0.14 is chosen. This fraction is indicated by crosses on top of the graphs in Fig.~\ref{fig:wf}.

\section{Results}
\label{sec:Results}

\begin{figure}
\centering
\includegraphics[width=0.99\linewidth]{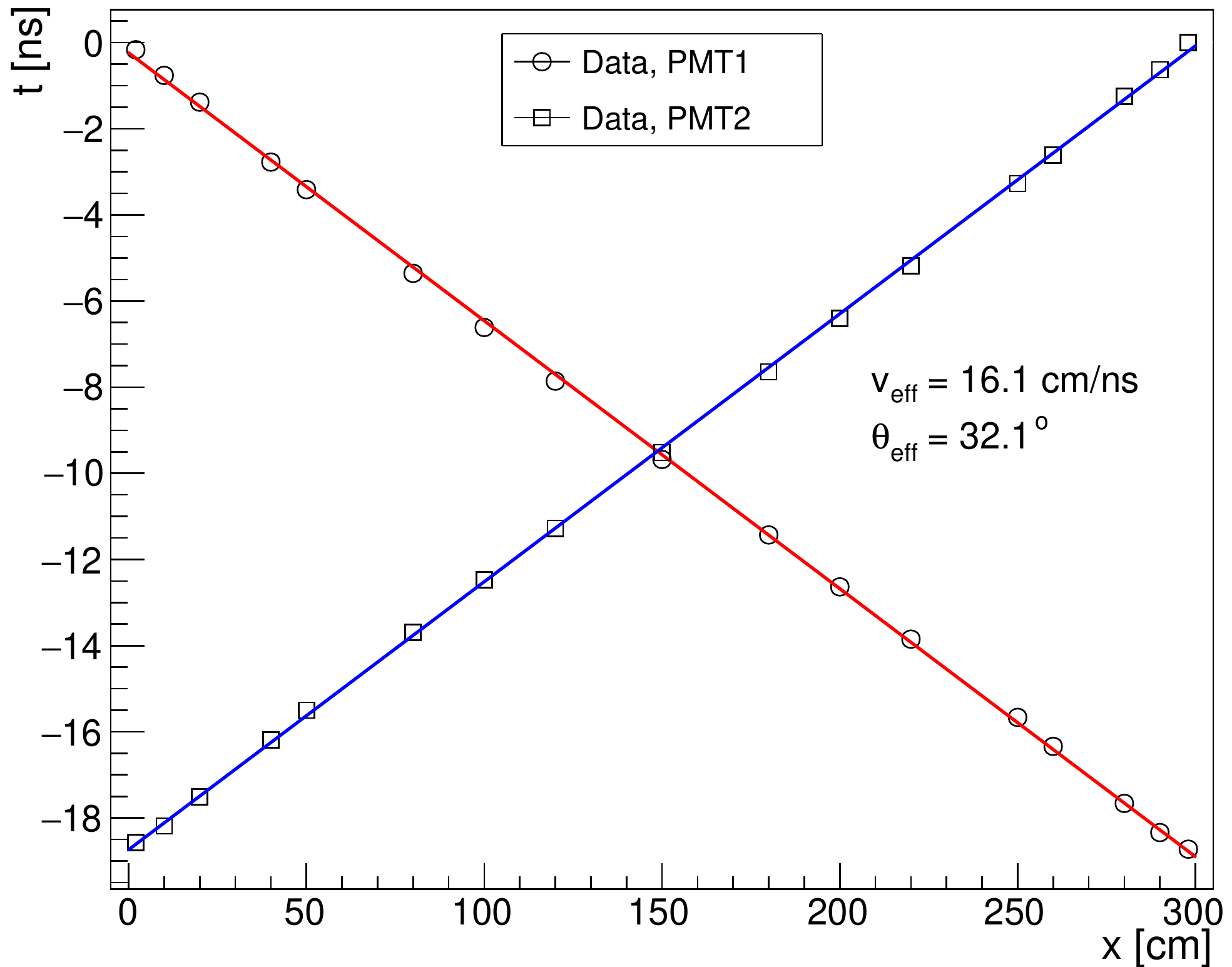}
\caption{Time as measured by PMT1 and PMT2 as a function of the interaction position $x$ along the bar.
}
\label{fig:TvsX}
\end{figure}

The dependency of the measured time versus position of the crossing point along the bar as viewed by both PMTs is shown in Fig.~\ref{fig:TvsX}. The graphs are approximated by linear functions whose slopes represent the effective average speed of light along the $x$ axis, which is found to be $v_{eff}=16.1$~cm/ns. Using the refraction index of the plastic one can convert this value into the effective average reflection angle: $\theta_{eff}=32.1^{\circ}$.

\begin{figure}
\centering
\includegraphics[width=0.97\linewidth]{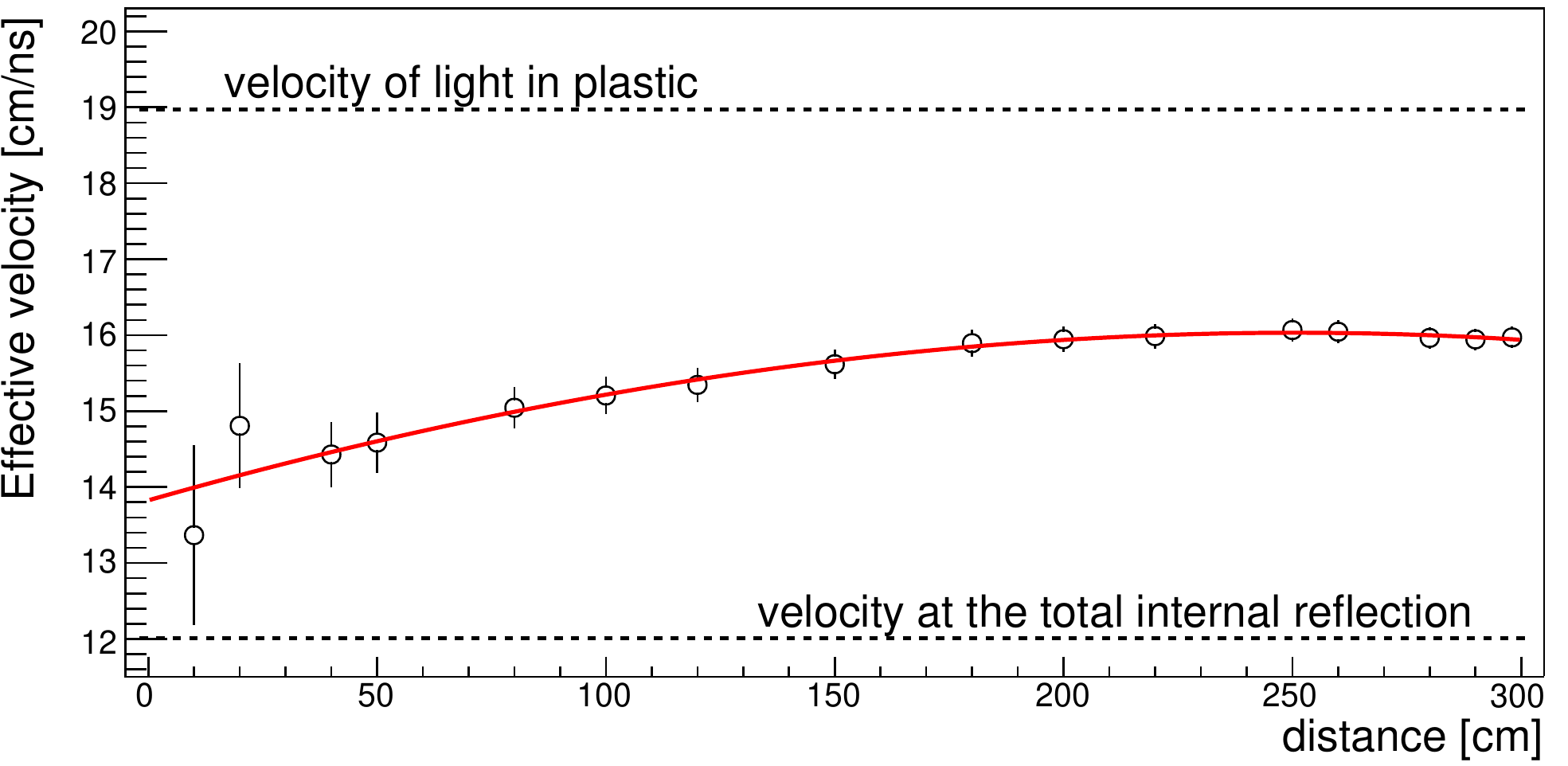}
\includegraphics[width=0.97\linewidth]{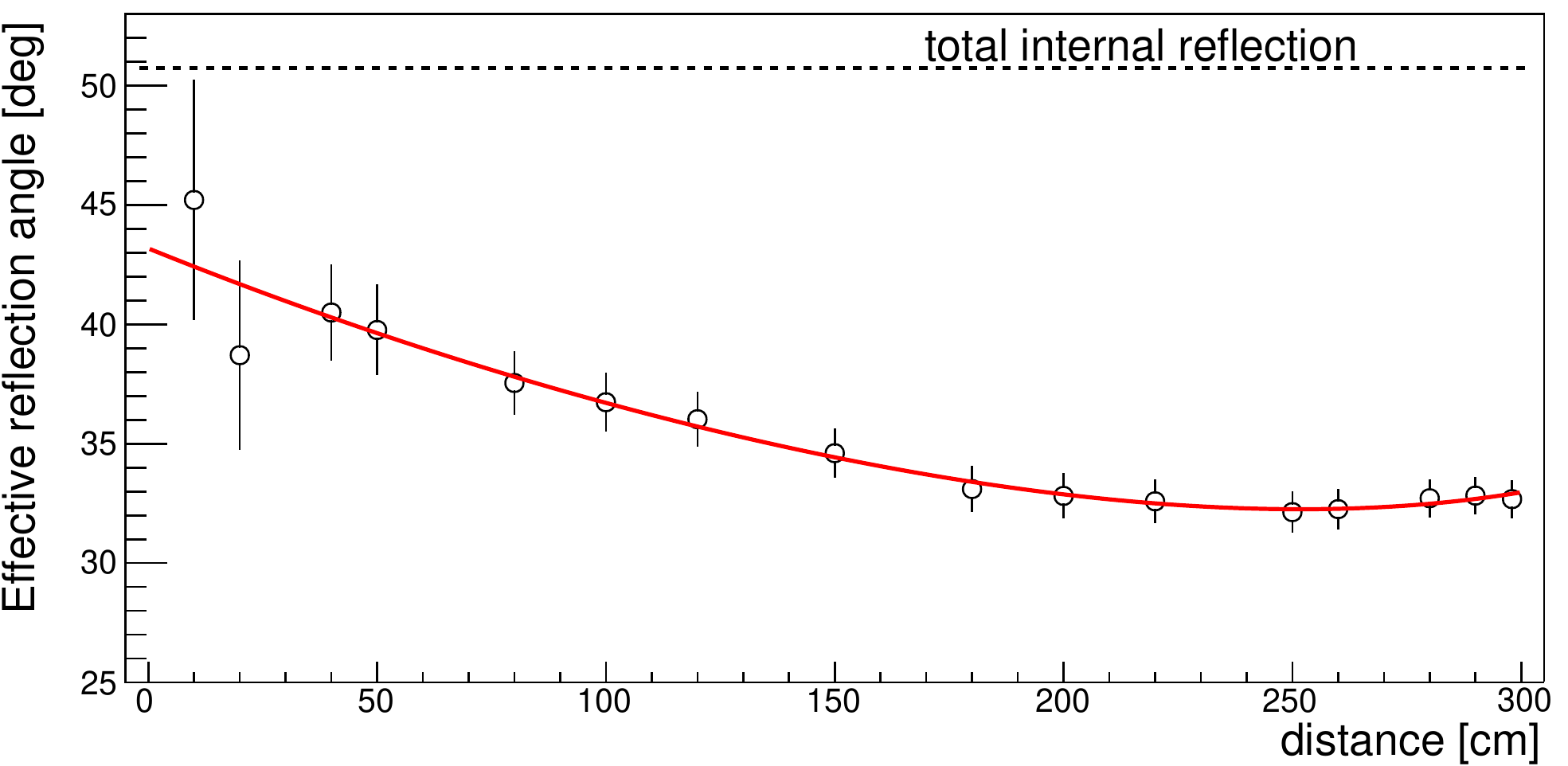}
\caption{The effective average signal velocity along the $x$ axis (top) and the corresponding reflection angle (bottom) as a function of position along the bar, as viewed by PMT1.}
\label{fig:VvsX}
\end{figure}

As one can see in Fig.~\ref{fig:TvsX}, points lie above the lines for interactions taking place nearby a PMT ($t\approx0$~ns) and they go below for interactions which are closer to the bar center. In order to understand this behavior we calculate $v_{eff}$ and $\theta_{eff}$ as a function of distance. Results are shown in Fig.~\ref{fig:VvsX}. The value of $v_{eff}$ in the proximity of the PMT is 14~ns/cm ($\theta_{eff}=43^{\circ}$).
It increases up to 16~ns/cm ($\theta_{eff}=32^{\circ}$) at $x=180$~cm and stays 
rather unchanged until the end of the bar.
Presumably this effect can be explained by the fact that photons at larger angles with respect to the surface of the bar have to travel longer distances before reaching the PMT,
and are therefore
more strongly affected by attenuation.


\begin{figure}
\centering
\includegraphics[width=0.97\linewidth]{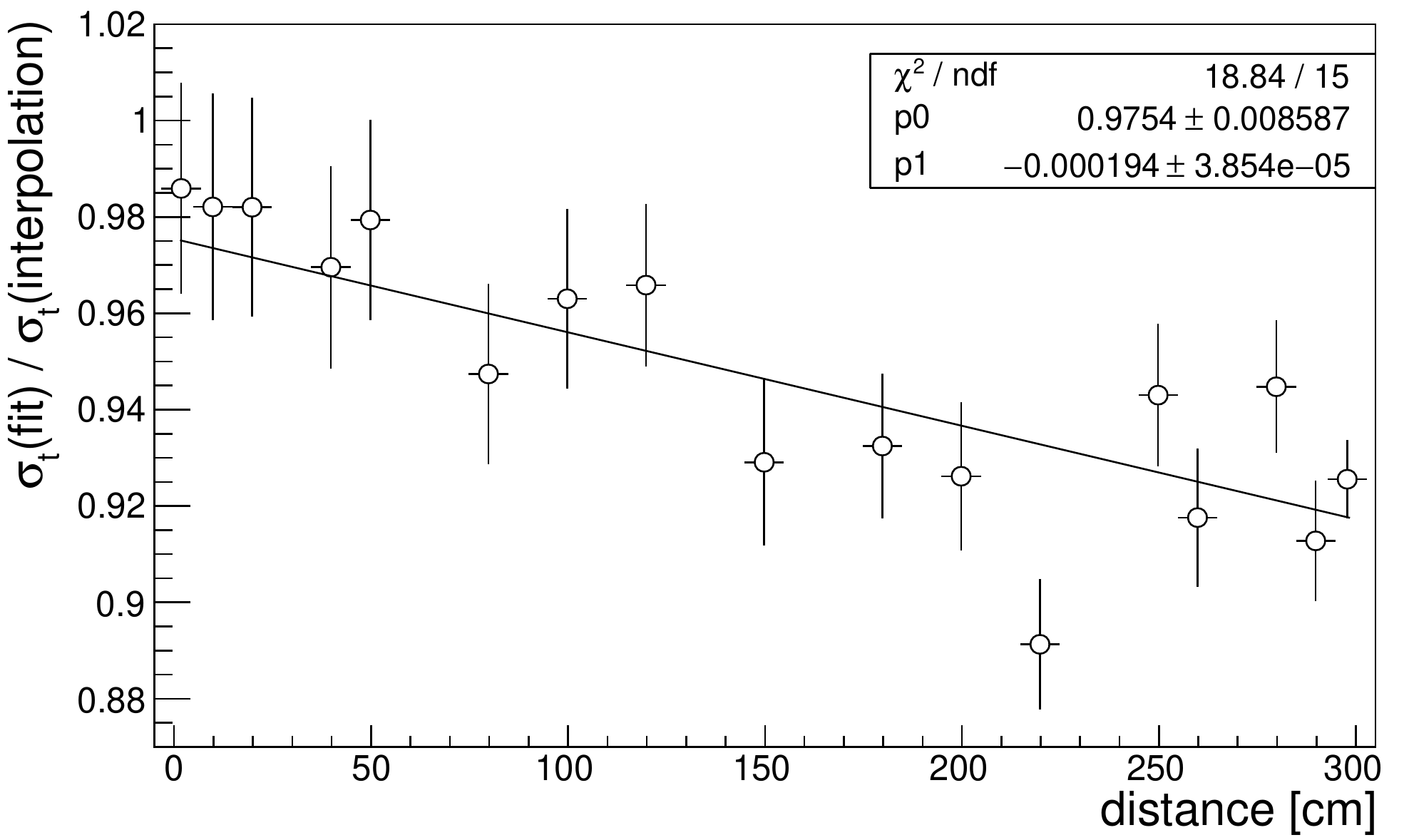}
\caption{Ratio of the time resolution obtained with the fit method and 
  that obtained using a linear interpolation in the dCFD technique.
  Data were recorded by a single PMT. 
  The line represents a fit by a linear function.}
\label{fig:FitvsInt}
\end{figure}


\begin{figure*}
\includegraphics[width=0.48\linewidth]{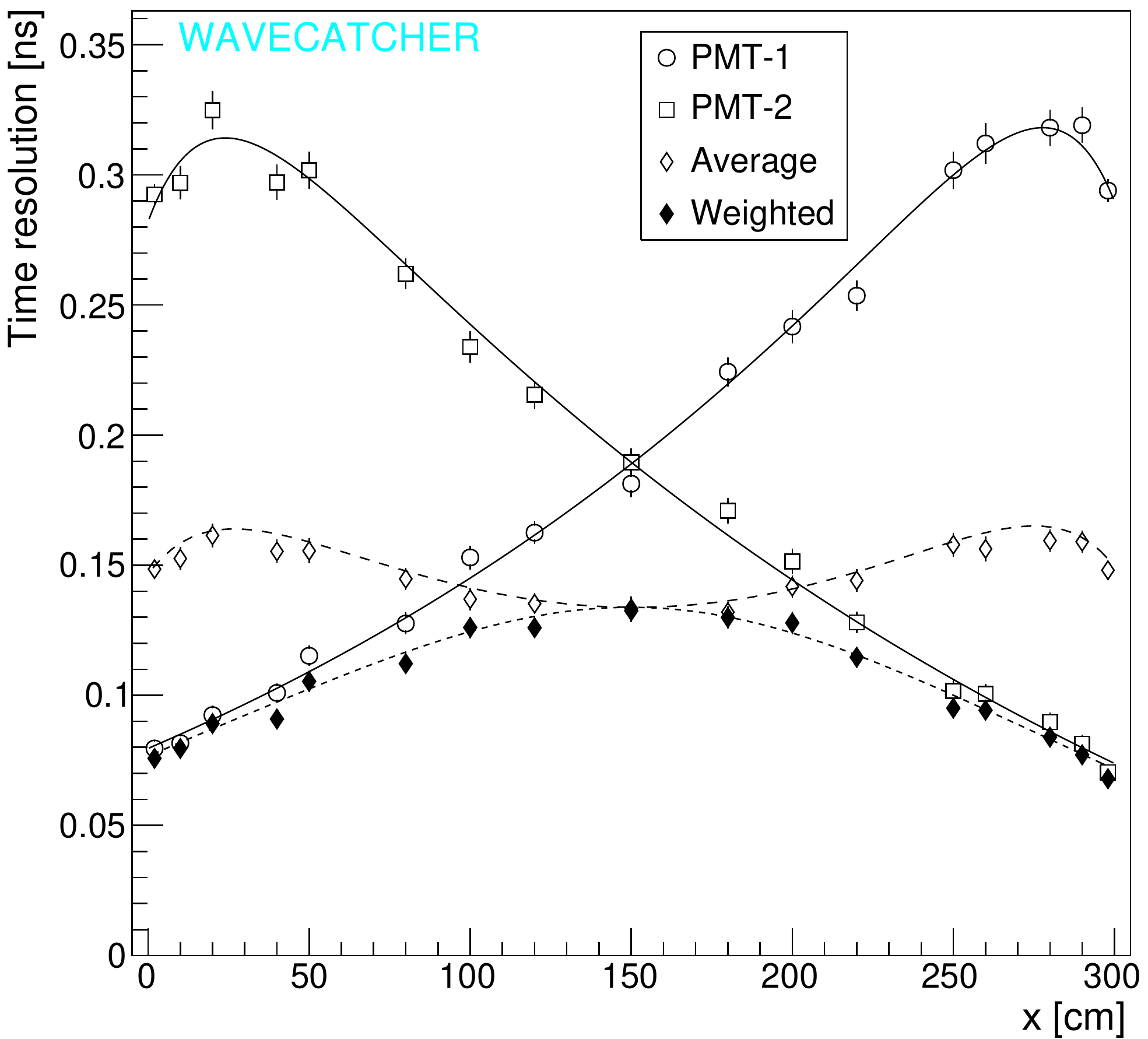}
\hfill
\includegraphics[width=0.48\linewidth]{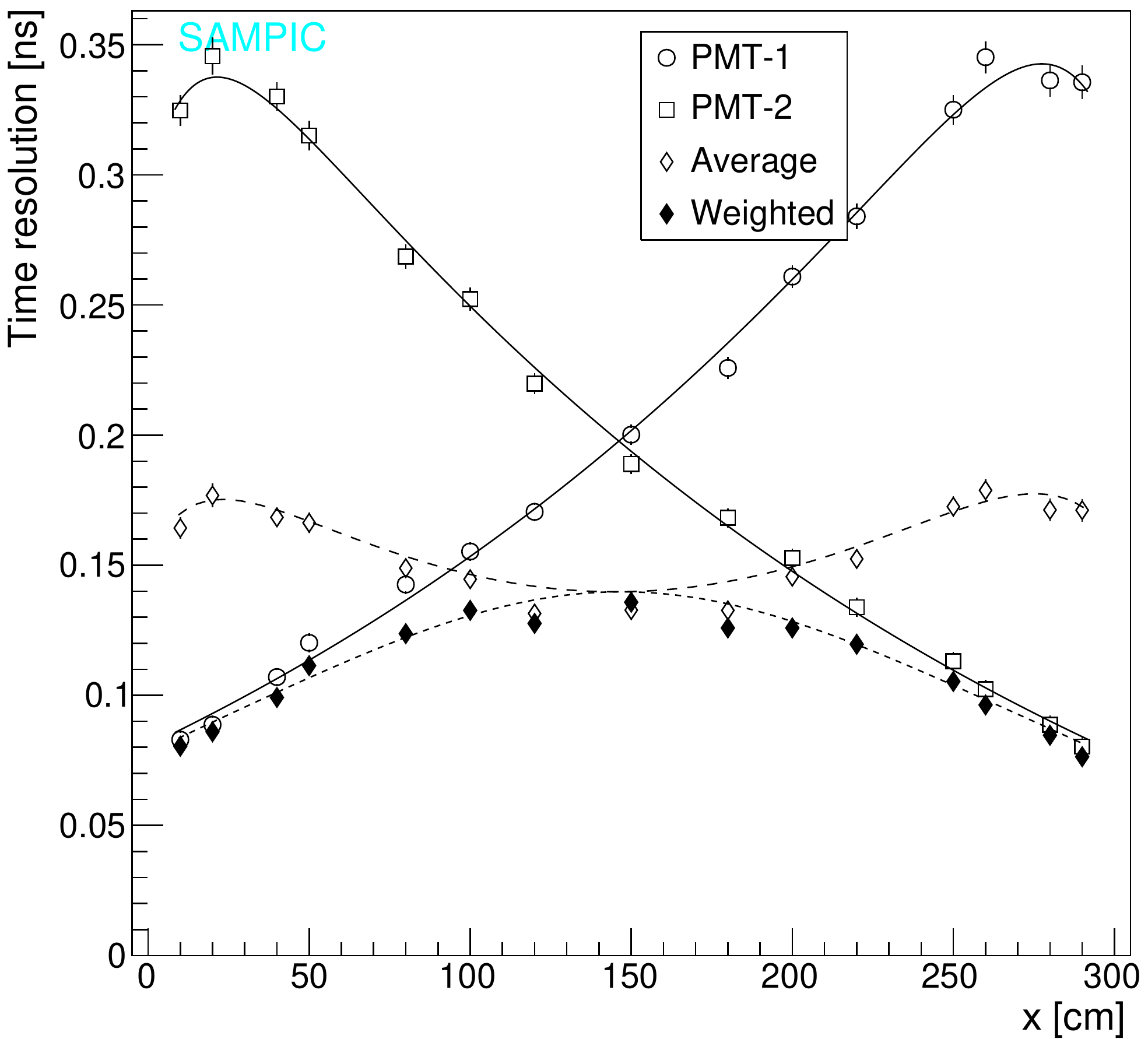}
\caption{Time resolution as a function of a distance obtained in measurements
  with WAVECATCHER (left) and SAMPIC (right),
  resulting from the dCFD analysis 
  with a fit of the signal rise edge.
  The bar resolution obtained using a simple average and 
  a weighted average of the signals from the two PMTs are also shown.
  Solid curves represent results of the empirical fit. 
  They are used to calculate parameters of the dashed curves 
  which correspond to the average resolutions.
}
\label{fig:dT_WC_SAM}
\end{figure*}


The time resolution as a function of longitudinal distance at $y=0$~cm obtained in measurements with WAVECATCHER is shown in the left panel of Fig.~\ref{fig:dT_WC_SAM}. The time of dCFD was extracted from the fit to the waveform as described in the previous section. The advantage of this method is demonstrated in Fig.~\ref{fig:FitvsInt} which shows the ratio between the resolution obtained with the fit method and the resolution obtained using a linear interpolation. The former provides better precision by 2\% in the proximity of the PMT, and the improvement reaches 8\% for the case of an interaction taking place at the far end of the bar. 

The time resolution of an individual PMT evolves from 80~ps for the crossing point near the phototube to 320~ps for the light propagation along the 280~cm distance.
A slight improvement of the resolution is observed in case of the crossing point being at the proximity of 300~cm. 
This could possibly be an effect of light reflected backwards. A similar effect was observed in Ref.\,\cite{Tsujita:1996tk}. The distribution is approximated by an analytic function (sum of two exponential functions and a constant) which is shown by a solid curve in Fig.~\ref{fig:dT_WC_SAM}. 

Measurements of time done 
by the two PMTs on both ends of the bar can be combined in two different ways. In the case where the position of the crossing point between the particle trajectory and the bar is unknown, the time of the interaction can only be calculated as a simple average
\begin{equation}
t_{aver} = \frac{t_1 + t_2}{2}
\label{Eq:aver}
\end{equation}
where $t_1$ and $t_2$ are the times measured by PMT1 and PMT2, respectively. 
On the other hand,
in a real experiment the crossing point can be well determined by other precision tracking detectors. In our case the interaction point is defined by the position of the reference trigger counters. Therefore one can weight the measurements of PMT1 and PMT2 according to their uncertainties $\sigma_{1}$ and $\sigma_{2}$
\begin{equation}
t_{weight} = \frac{t_1/\sigma_{1}^2 + t_2/\sigma_{2}^2}{1/\sigma_{1}^2 + 1/\sigma_{2}^2}
\label{Eq:weight}
\end{equation}
The time resolutions extracted from fits of the distributions obtained with Eq.~(\ref{Eq:aver}) and Eq.~(\ref{Eq:weight}) are shown in Fig.~\ref{fig:dT_WC_SAM} by open and full diamond symbols, respectively. Both approaches provide 133~ps resolution in the central region of the bar. On the other hand, in the outer parts of the bar the phototube which is closer to the crossing point makes possible a substantially better resolution. In this region the weighted average provides a significant advantage, with a time resolution of 80~ps, while the resolution of the simple average is around 160~ps. Therefore, the time resolution of the bar can be regarded as being 150~ps for the simple average and 100~ps for the weighted average approach over the entire bar length of 3~m.

\begin{figure}
\centering
\includegraphics[width=0.97\linewidth]{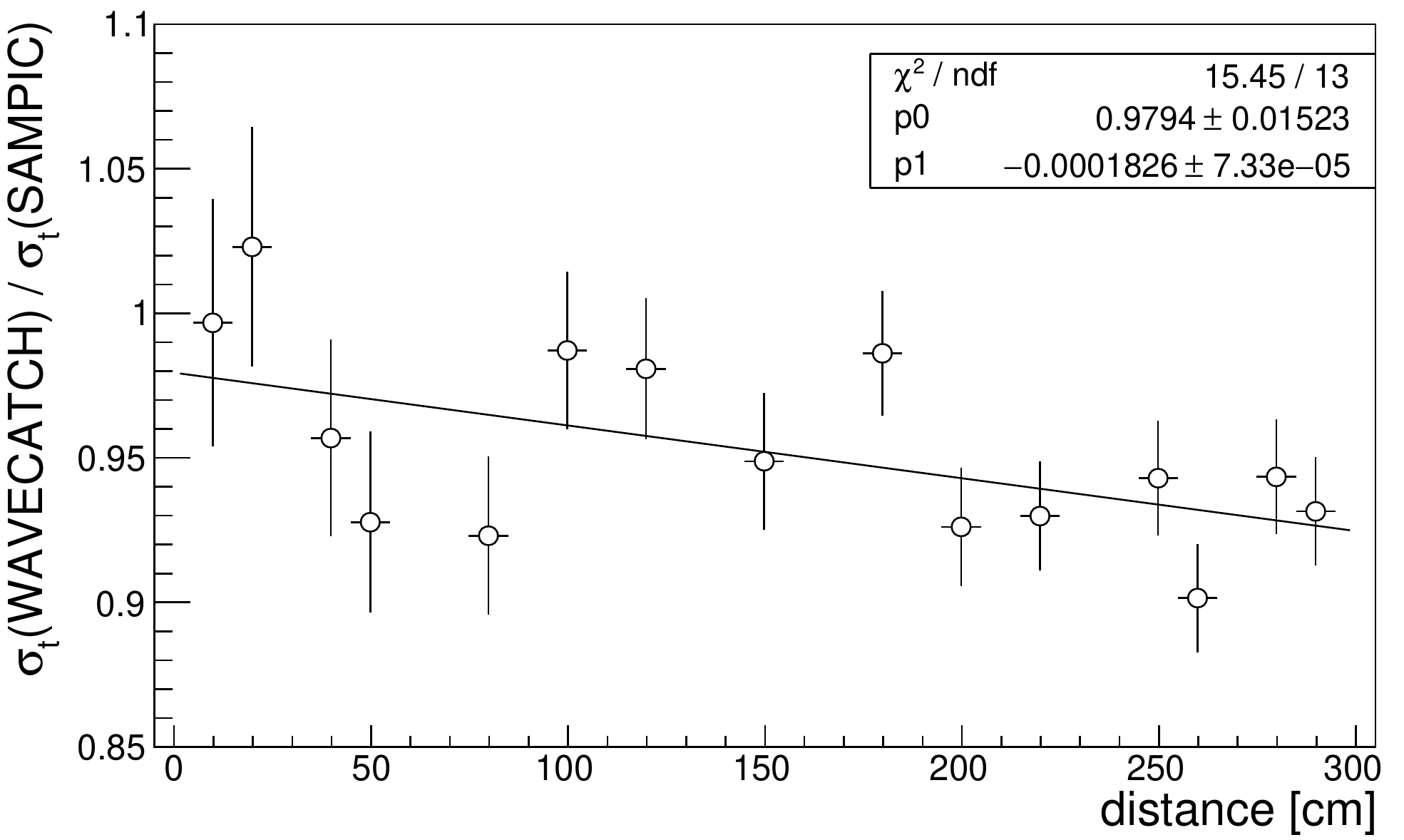}
\caption{Ratio of the time resolution obtained  in the dCFD analysis 
  of a single PMT with data acquired by WAVECATCHER and with SAMPIC.
  The line represents a fit by a linear function.}
\label{fig:WCvsSAM}
\end{figure}

The time resolution as a function of longitudinal distance at $y=0$~cm obtained in measurements with SAMPIC is shown in the right panel of Fig.~\ref{fig:dT_WC_SAM}. The time resolution of the bar when measured by a single PMT varies from 80~ps to 340~ps. The resolution in the center of the bar, when calculated as an average for two PMTs, is found to be 140~ps. The slight differences in the values obtained with SAMPIC and with WAVECATCHER are mostly a consequence of a shorter waveform recorded by SAMPIC. 
The ratio of the two is shown in Fig.~\ref{fig:WCvsSAM}.


\subsection{Resolution vs transverse coordinate}

\begin{figure}[h]
\centering
\includegraphics[width=0.99\linewidth]{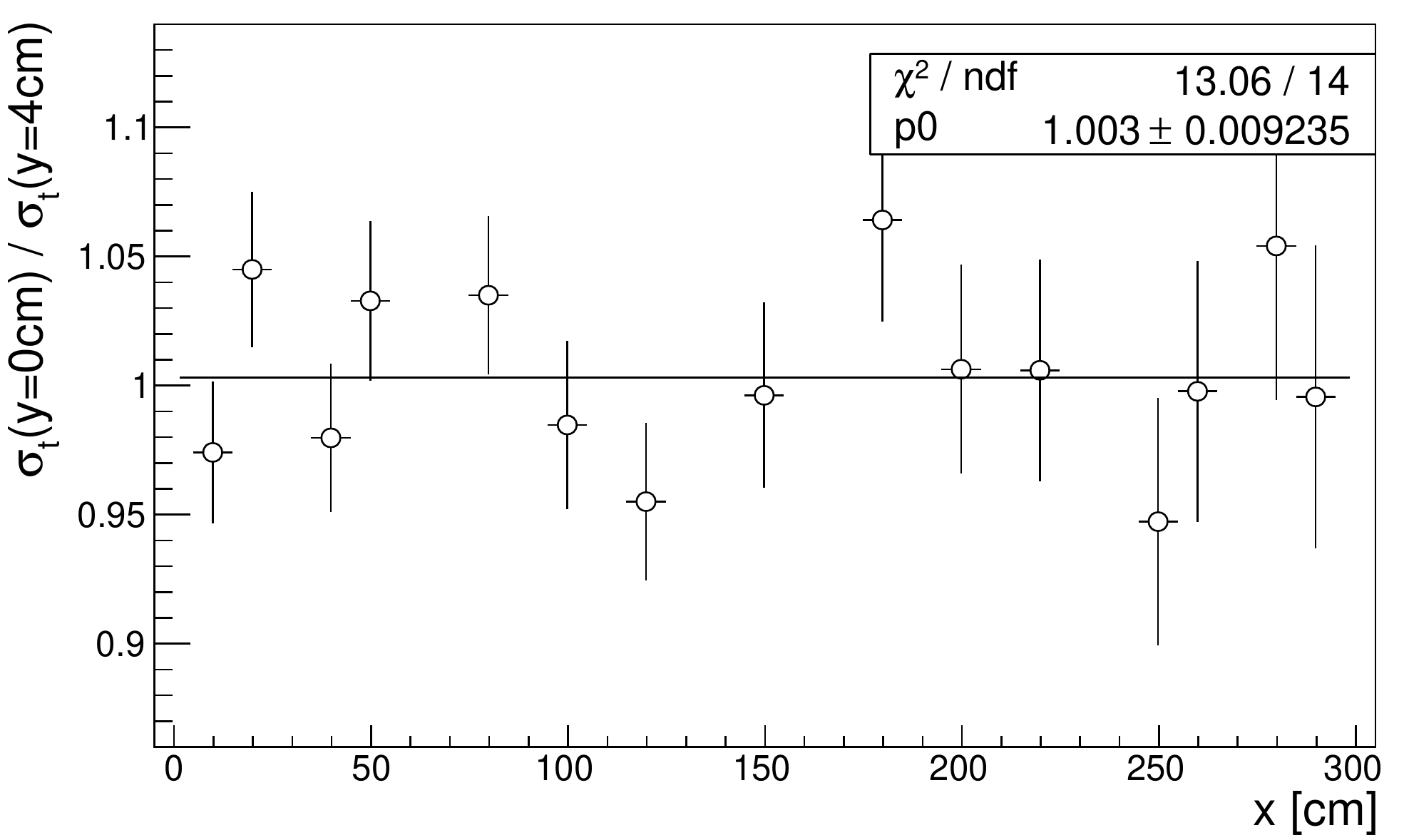}
\caption{Ratio of the time resolutions obtained in scans at $y=0$~cm
  and $y=4$~cm for PMT2. The line represents a fit by a constant function.
}
\label{fig:dTvsY}
\end{figure}

The time resolution of the bar was studied as a function of $y$, using a second scan with the position of the bar shifted by $4$~cm upward. The ratio of the time resolutions obtained in scans at $y=0$~cm and $y=4$~cm is shown in Fig.~\ref{fig:dTvsY}. A fit with a constant function gives a value compatible with 1, showing that the shift does not alter the time 
resolution.


\subsection{Phenomenological analysis}
\label{Sec:phenom_analys}

In order to quantify the effects discussed at the beginning of this section, a phenomenological analysis is performed to describe the resolution $\sigma_t$ in terms of physics-motivated contributions.

The time resolution as a function of the light propagation distance $l$ can be decomposed as~\cite{Kuhlen:1990ny,Perrino:1996pr}
\begin{equation}
\sigma_t(l) = \sqrt{ \frac{\sigma^2_{sci+PMT}}{N(l)} + \frac{(\sigma_{length}\cdot l)^2}{N(l)} + \sigma_{el}^2}
\label{Eq:sigma_t}
\end{equation}
where the parameter $\sigma_{sci+PMT}$ represents a contribution from a spread of the photon emission time of the scintillator and the time jitter of the PMT; $\sigma_{length}$ accounts for a time spread due to the light transmission; and $\sigma_{el}$ is the contribution which is independent of the light strength, such as a noise of the readout electronics.
The number of observed photoelectrons $N$ depends on distance and can be directly extracted in the experiment.

\begin{figure}[h]
\centering
\includegraphics[width=0.99\linewidth]{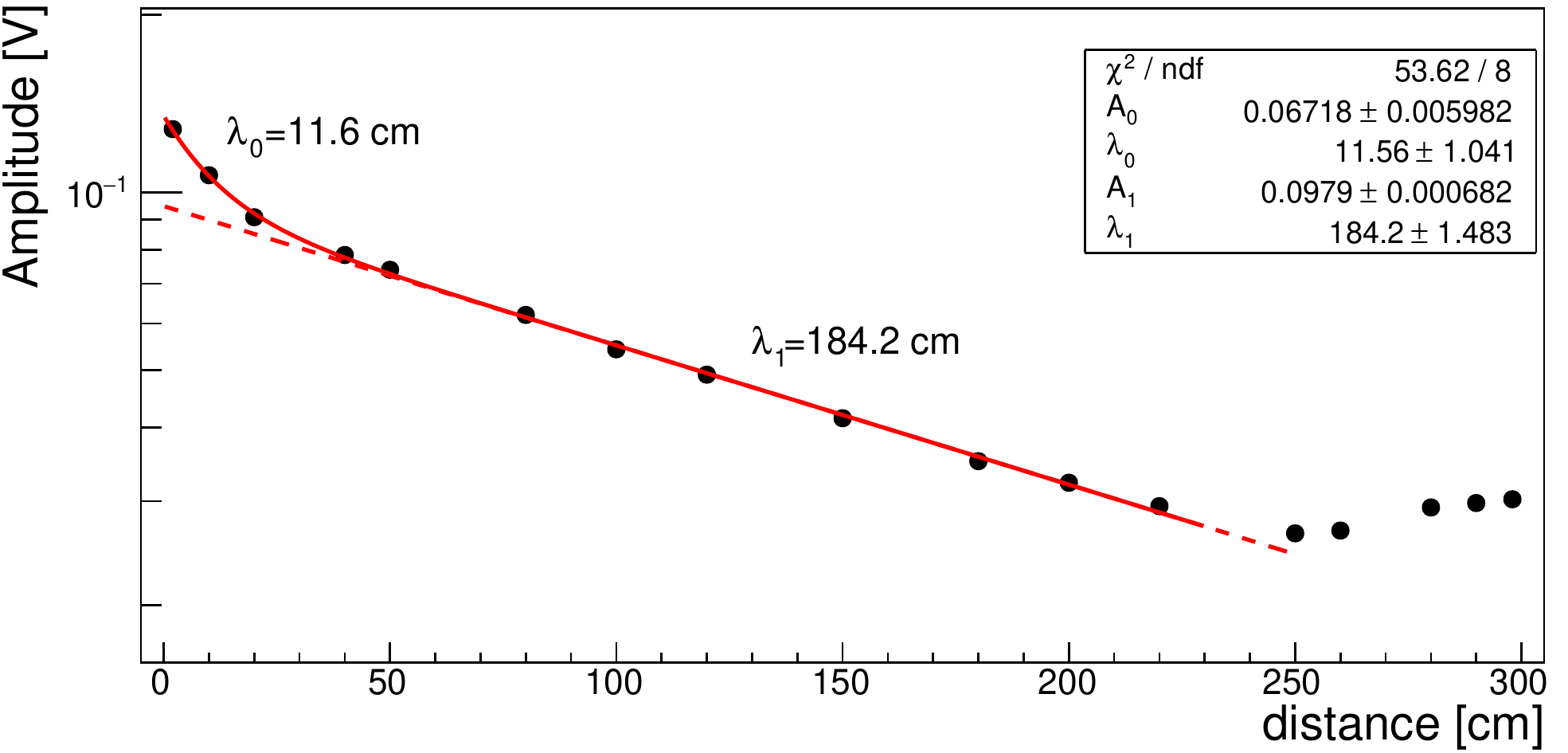}
\caption{
  The most probable value of the pulse height spectrum
  versus the light propagation distance.
  }
\label{fig:FitN}
\vspace{0.5cm}
\centering
\includegraphics[width=0.99\linewidth]{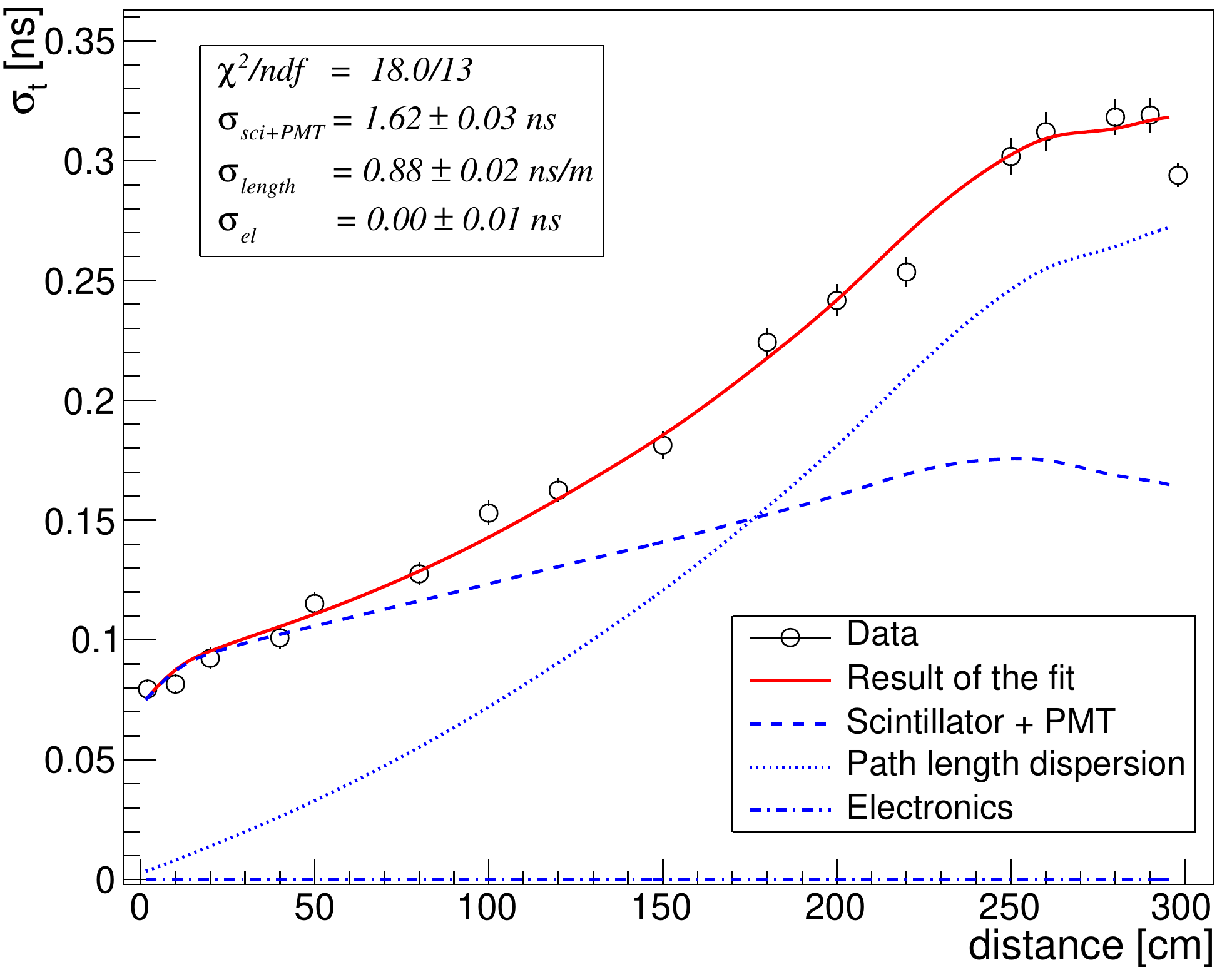}
\caption{Time resolution as a function of distance. The circles represent data and the solid curve shows the result of the phenomenological fit from Eq.\,(\ref{Eq:sigma_t}). Different contributions are shown as lines of different styles.}
\label{fig:FitT}
\end{figure}

The most probable value of the pulse height spectrum
is shown in Fig.~\ref{fig:FitN}
as a function of the light propagation distance.
One can distinguish several regions.
The distribution falls down more steeply for the first 20~cm.
It could be explained by a strong dependence of the attenuation length on 
the photon wavelength as shown in Fig.~\ref{fig:sensitivity}, 
i.e. the contribution of short-wavelength photons is significant
for the short propagation of light and rapidly diminishes at larger distances.
The behavior of the light transmission at larger distance is mainly dictated by photons which undergo total internal reflection, traveling therefore a much longer distance inside the plastic. For total reflection, the angle between the photon trajectory and the 
surface of the bar has to be smaller than 
$90^\circ - {\rm asin}(1/n) = 51^\circ$, where $n=1.58$ is the refraction index of the plastic.
When fitting by a sum of two exponents one obtains the effective attenuation length $\lambda = 184.2$~cm for this region. For the last 50~cm of the bar the attenuation is not observed.


In order to estimate the number of photoelectrons collected by phototubes the procedure described in Ref.\,\cite{Tsujita:1996tk} is applied. The number of detected photoelectrons is assumed to be proportional to the signal amplitude. The spread of the asymmetry of amplitudes of PMT1 and PMT2 signals provides a lower estimate for the number of photoelectrons. The procedure results in $N_1 = N_2 = 132$ at the center of the bar. This value is used to rescale the amplitude distribution to that of the number of photoelectrons which, in turn, is used in the fit of the time resolution $\sigma_t (l)$ with the function of Eq.~(\ref{Eq:sigma_t}).


The distribution of the time resolution for PMT1 overlaid with the fitting function is shown in Fig.~\ref{fig:FitT}. At short distance the resolution is driven by the photoemission properties of the plastic. 
The PMT also contributes to this uncertainty via the transit time spread but at a much lower level. 
The value obtained from the fit is $\sigma_{sci+PMT}=1.62 \pm 0.03$~ns. In the middle of the bar the contribution of the path length dispersion becomes equal in size with 
the photoemission spread.
This contribution is linear in distance and it enlarges the uncertainty as $\sigma_{length}=0.88 \pm 0.02$~ns/m. The last term which reflects the contribution from the readout electronics, $\sigma_{el}$, is not significant. The fitting function is able to approximate reasonably the time resolution except as the point at the far end of the bar. This improvement of the resolution is likely to be due to the presence of reflected light  and is not described by Eq.\,(\ref{Eq:sigma_t}).


\subsection{Resolution vs incident angle}

\begin{figure}
\centering
\includegraphics[width=0.99\linewidth]{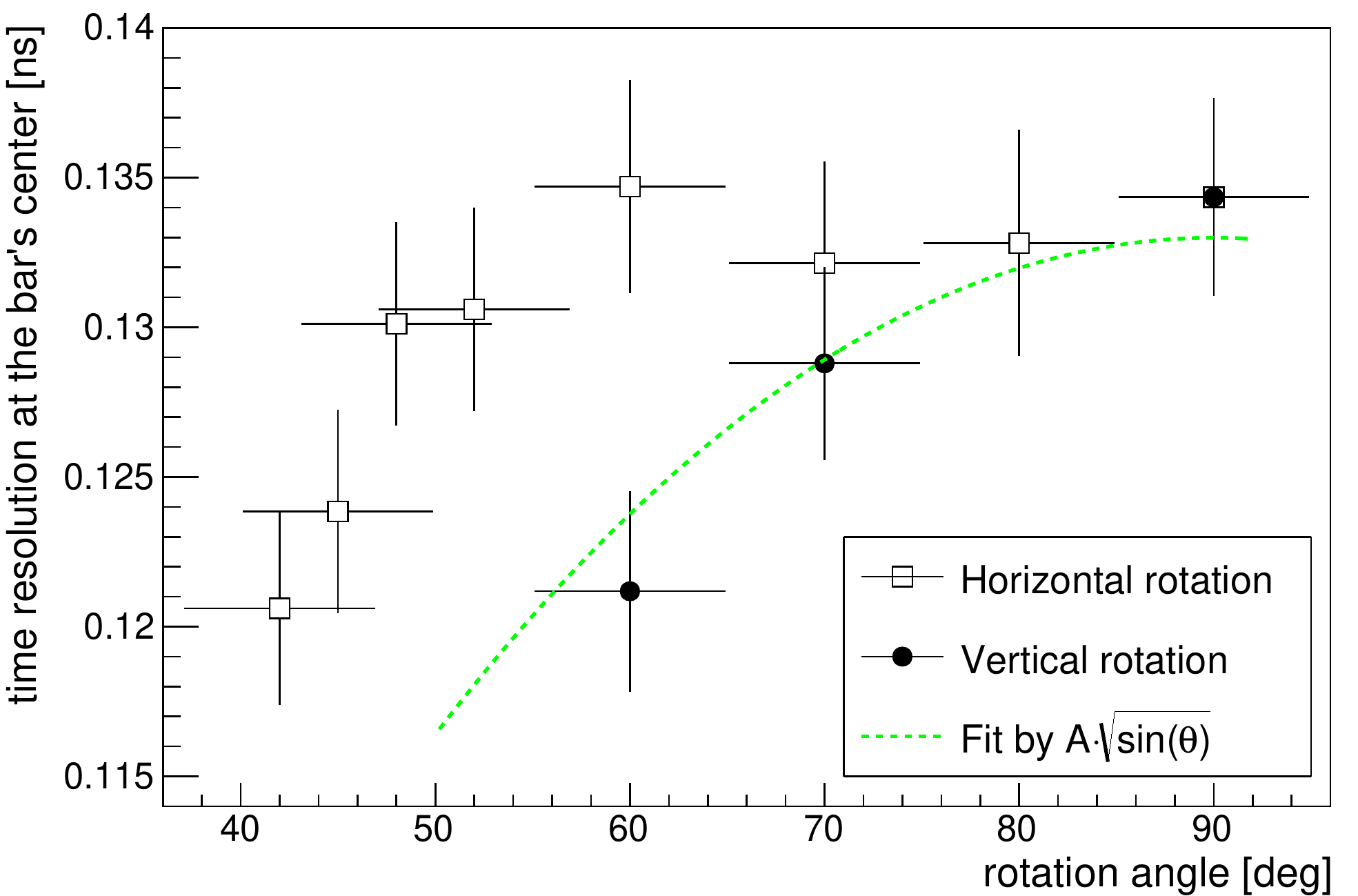}
\caption{Time resolution at the center of the bar ($x=150$~cm) for 
  various angles of incidence with respect to the bar surface. 
  Circles correspond to a bar rotation in the vertical $yz$ plane.
  The dashed line is a fit assuming that the number of photons 
  from the particle path length inside the plastic is the only contribution to the resolution.
  Square markers represent measurements with a rotation in the horizontal $xz$ plane. 
}
\label{fig:dT_rot}
\end{figure}

The behavior of $\sigma_t$ as a function of the angle between the surface of the bar and the beam trajectory is also studied. To do so the bar was rotated in either the horizontal $xz$ or the vertical $yz$ plane. The time was calculated as an average between measurements of the two PMTs at the center of the bar. Results are presented in Fig.~\ref{fig:dT_rot}.

There are two effects which can be discussed in this respect. The track length inside the bar gets longer in case of the oblique incidence, thus the number of emitted photons increases proportionally and the time resolution improves as a square root of this number. Indeed the measurements follow a $\sqrt{\sin\phi}$ behavior for the rotation in the vertical plane. In case of the rotation in the horizontal plane, an additional uncertainty comes from the fact that the entrance and exit points of a track are located at different distances from the PMT.
It increases the rise time of a signal, thus worsening the time resolution. As it can be seen in Fig.~\ref{fig:dT_rot}, the resolution does not change in the interval from 50$^\circ$ to 90$^\circ$, which presumably indicates a compensation between the increase of the number of photons and the smearing of the signal rising edge. This observation is in general agreement with results presented in Refs~\cite{Tsujita:1996tk,Wu:2005xk}. The resolution improves for angles smaller than 50$^\circ$. 
In this region a measurable fraction of Cherenkov photons can reach one of PMTs without reflections. 
The number of Cherenkov photons increases for smaller values of angle
which presumably explains the improved time resolution.

\section{Conclusions}
\label{sec:Conclusions}

The timing properties of a plastic scintillator counter with dimensions 300~cm $\times$ 11~cm $\times$ 2.5~cm were studied in detail using the test-beam facility of the East Area of the CERN PS. 
Two waveform digitizer modules, WAVECATCHER and SAMPIC, were exploited for the data acquisition.

The time resolution of the counter when measured by a single PMT varies from 80~ps to 320~ps and from 80~ps to 340~ps in case of WAVECATCHER and SAMPIC, respectively. The resolution in the center of the bar, when calculated as an average of the two PMTs, is found to be 133~ps and 140~ps for the two DAQ modules, respectively. The behavior of the resolution as a function of the light propagation distance was successfully parametrized with a physics-motivated model. 
Measurements along the central line of the bar show the same time resolution as those displaced by 4~cm from the central line.
%
The behavior of the time resolution versus incident angle for rotations in horizontal and vertical planes was also discussed and understood.

These results confirm that 
the use of long scintillator counters can provide a time resolution of approximately 100~ps for a large-scale detector used in particle physics experiments. The SAMPIC waveform digitizer appears to be adequate for such applications, especially if a self-triggering capacity and a tolerance for high signal rates is desired.


\section*{Acknowledgments}

This work was supported by the Swiss National Science Foundation. 
We would also like to acknowledge the contribution of FAST (COST action TD1401) 
for inspiring a collaboration between the engineering group and the researchers.
We thank the European Organization for Nuclear Research for support and hospitality 
and, in particular, the operating crews of the CERN PS accelerator and 
beamlines who made the measurements possible.


\bibliography{article}

\begin{thebibliography}{10}
\expandafter\ifx\csname url\endcsname\relax
  \def\url#1{\texttt{#1}}\fi
\expandafter\ifx\csname urlprefix\endcsname\relax\def\urlprefix{URL }\fi
\expandafter\ifx\csname href\endcsname\relax
  \def\href#1#2{#2} \def\path#1{#1}\fi

\bibitem{Bertoni:2010by}
R.~Bertoni, et~al., {The design and commissioning of the MICE upstream
  time-of-flight system}, Nucl. Instrum. Meth. A615 (2010) 14--26.
\newblock \href {http://arxiv.org/abs/1001.4426} {\path{arXiv:1001.4426}},
  \href {http://dx.doi.org/10.1016/j.nima.2009.12.065}
  {\path{doi:10.1016/j.nima.2009.12.065}}.

\bibitem{Ahmet:1990eg}
K.~Ahmet, et~al., {The OPAL detector at LEP}, Nucl. Instrum. Meth. A305 (1991)
  275--319.
\newblock \href {http://dx.doi.org/10.1016/0168-9002(91)90547-4}
  {\path{doi:10.1016/0168-9002(91)90547-4}}.

\bibitem{Perrino:1996pr}
R.~Perrino, et~al., {Timing measurements in long rods of BC408 scintillators
  with small cross-sectional sizes}, Nucl. Instrum. Meth. A381 (1996) 324--329.
\newblock \href {http://dx.doi.org/10.1016/S0168-9002(96)00812-1}
  {\path{doi:10.1016/S0168-9002(96)00812-1}}.

\bibitem{Tsujita:1996tk}
T.~Tsujita, Y.~Asano, H.~Hamasaki, S.~Mori, K.~Yusa, R.~D. Kephart, {Test of a
  3 m long, 4x4 cm$^2$ time-of-flight (TOF) scintillation counter using 38x38
  mm$^2$ fine-mesh photomultipliers in magnetic fields up to 1.5 T}, Nucl.
  Instrum. Meth. A383 (1996) 413--423.
\newblock \href {http://dx.doi.org/10.1016/S0168-9002(96)00871-6}
  {\path{doi:10.1016/S0168-9002(96)00871-6}}.

\bibitem{Denisov:2004ag}
S.~Denisov, A.~Dzierba, R.~Heinz, A.~Klimenko, V.~Samoilenko, E.~Scott,
  P.~Smith, S.~Teige, {Systematic studies of timing characteristics for 2 m
  long scintillation counters}, Nucl. Instrum. Meth. A525 (2004) 183--187.
\newblock \href {http://dx.doi.org/10.1016/j.nima.2004.03.043}
  {\path{doi:10.1016/j.nima.2004.03.043}}.

\bibitem{Kichimi:2000}
H.~Kichimi, et~al., {The BELLE TOF system}, Nucl. Instrum. Meth. A453 (2000)
  315--320.

\bibitem{Wu:2005xk}
C.~Wu, et~al., {The timing properties of a plastic time-of-flight scintillator
  from a beam test}, Nucl. Instrum. Meth. A555 (2005) 142--147.
\newblock \href {http://dx.doi.org/10.1016/j.nima.2005.09.029}
  {\path{doi:10.1016/j.nima.2005.09.029}}.

\bibitem{Bernet:2005yy}
C.~Bernet, et~al., {The COMPASS trigger system for muon scattering}, Nucl.
  Instrum. Meth. A550 (2005) 217--240.
\newblock \href {http://dx.doi.org/10.1016/j.nima.2005.05.043}
  {\path{doi:10.1016/j.nima.2005.05.043}}.

\bibitem{Abgrall:2014xwa}
N.~Abgrall, et~al., {NA61/SHINE facility at the CERN SPS: beams and detector
  system}, JINST 9 (2014) P06005.
\newblock \href {http://arxiv.org/abs/1401.4699} {\path{arXiv:1401.4699}},
  \href {http://dx.doi.org/10.1088/1748-0221/9/06/P06005}
  {\path{doi:10.1088/1748-0221/9/06/P06005}}.

\bibitem{Bonivento:2013jag}
W.~Bonivento, et~al., {Proposal to Search for Heavy Neutral Leptons at the SPS
  }\href {http://arxiv.org/abs/1310.1762} {\path{arXiv:1310.1762}}.

\bibitem{Anelli:2015pba}
M.~Anelli, et~al., {A facility to Search for Hidden Particles (SHiP) at the
  CERN SPS }\href {http://arxiv.org/abs/1504.04956} {\path{arXiv:1504.04956}}.

\bibitem{Delagnes:WAVECATCHER}
D.~Breton, E.~Delagnes, J.~Maalmi, P.~Rusquart,
  \href{http://hal.in2p3.fr/in2p3-00995691}{{The WaveCatcher Family of
  SCA-Based 12-Bit 3.2-GS/s Fast Digitizers}}, {RT2014 - 19th Real-Time
  Conference}, poster (May 2014).
\newline\urlprefix\url{http://hal.in2p3.fr/in2p3-00995691}

\bibitem{Delagnes:SAMPIC}
E.~Delagnes, D.~Breton, H.~Grabas, J.~Maalmi, P.~Rusquart, M.~Saimpert,
  \href{http://hal.in2p3.fr/in2p3-01082061}{{The SAMPIC Waveform and Time to
  Digital Converter}}, in: {2014 IEEE Nuclear Science Symposium and Medical
  Imaging Conference (2014 NSS/MIC)}, Seattle, United States, 2014, sce
  Electronique.
\newline\urlprefix\url{http://hal.in2p3.fr/in2p3-01082061}

\bibitem{SCIONIX}
SCIONIX HOLLAND BV, Radiation Detectors \& Crystals, http://scionix.nl.

\bibitem{Hamamatsu}
Hamamatsu, http://www.hamamatsu.com.

\bibitem{Brooks_CLASS}
W.~Brooks, {A simple model for the propagation of internally reflected light
  through a scintillator and light guide system}, CLAS-NOTE 94-017.

\bibitem{Nelson}
M.~Nelson, B.~Rooney, D.~Dinwiddie, B.~G., {Analysis of digital timing methods
  with BaF2 scintillators}, Nucl. Instrum. Meth. A505 (2003) 324--327.
\newblock \href {http://dx.doi.org/10.1016/S0168-9002(03)01078-7}
  {\path{doi:10.1016/S0168-9002(03)01078-7}}.

\bibitem{Wang:2015rva}
J.~Wang, S.~Liu, Q.~An, {Waveform Timing Performance of a 5 GS/s Fast Pulse
  Sampling Module with DRS4}, Chin. Phys. C39~(10) (2015) 106101.
\newblock \href {http://arxiv.org/abs/1501.00651} {\path{arXiv:1501.00651}},
  \href {http://dx.doi.org/10.1088/1674-1137/39/10/106101}
  {\path{doi:10.1088/1674-1137/39/10/106101}}.

\bibitem{Delagnes:2016hdo}
E.~Delagnes, {What is the theoretical time precision achievable using a dCFD
  algorithm? }\href {http://arxiv.org/abs/1606.05541}
  {\path{arXiv:1606.05541}}.

\bibitem{Kuhlen:1990ny}
M.~Kuhlen, M.~Moszynski, R.~Stroynowski, E.~Wicklund, B.~Milliken, {Timing
  properties of long scintillation counters based on scintillating fibers},
  Nucl. Instrum. Meth. A301 (1991) 223--229.
\newblock \href {http://dx.doi.org/10.1016/0168-9002(91)90463-Z}
  {\path{doi:10.1016/0168-9002(91)90463-Z}}.

\end{thebibliography}

\end{document}